\documentclass[12pt, a4paper]{article}
\pdfoutput=1

\usepackage[height=25cm,width=16cm]{geometry}
\usepackage{amsfonts}
\usepackage{amsmath}
\usepackage{amssymb}
\usepackage{ascmac}
\usepackage{dcolumn}
\usepackage{bm,here}
\usepackage{subfig}
\usepackage{comment}
\usepackage{ifpdf}
\usepackage{slashed}
\usepackage{colortbl}
\usepackage{color}
\usepackage[normalem]{ulem}
\usepackage[mathscr]{eucal}
\usepackage{cite}
\usepackage[sort&compress, numbers, merge]{natbib}
\usepackage[utf8]{inputenc}
\usepackage{cancel}

\ifpdf
  \usepackage{graphicx}     
  \usepackage[bookmarksopen,colorlinks=true,linkcolor=bblue,citecolor=ppink,urlcolor=ppink,linktocpage=false]{hyperref}
\else     
  \usepackage[dvipdfmx]{graphicx}     
  \usepackage[dvipdfmx,bookmarksopen,colorlinks=true,linkcolor=bblue,citecolor=ppink,urlcolor=ppink,linktocpage=false]{hyperref}
\fi

\usepackage{multicol}
\definecolor{red}{rgb}{1,0,0}
\definecolor{ppink}{rgb}{0.921545,0.440586,0.687243}
\definecolor{bblue}{rgb}{0.400000,0.400000,1.000000}
\usepackage[charter]{mathdesign}

\begin{document}

\begin{titlepage}

\begin{flushright}
  IPMU18-0003
\end{flushright}

\begin{center}

\vskip 3cm
{\Large \bf
	Direct Detection of Ultralight Dark Matter \\
	\vskip 0.2cm
	via Astronomical Ephemeris
}

\vskip 2cm
{\large
Hajime Fukuda,
Shigeki Matsumoto
and
Tsutomu T. Yanagida
}

\vskip 0.5cm
\begin{tabular}{l}
{\sl Kavli IPMU (WPI), UTIAS, University of Tokyo, Kashiwa, 277-8583, Japan}
\end{tabular}

\vskip 5.0cm
\begin{abstract}
\noindent
A novel idea of the direct detection to search for a ultralight dark matter based on the interaction between the dark matter and a nucleon is proposed. Solar system bodies feel the dark matter wind and it acts as a resistant force opposing their motions. The astronomical ephemeris of solar system bodies is so precise that it has a strong capability to detect a dark matter whose mass is much lighter than {\cal O}(1)\,eV. We have estimated the resistant force based on the calculation of the elastic scattering cross section between the dark matter and the bodies beyond the Born approximation, and show that the astronomical ephemeris indeed put a very strong constraint on the interaction between the dark matter and a nucleon, depending on how smoothly the ultralight dark matter is distributed at the scale smaller than the celestial bodies in our solar system.
\end{abstract}

\end{center}

\end{titlepage}

\newpage
\setcounter{page}{1}

\section{Introduction}
\label{sec: intro}

There is no doubt about the existence of dark matter in our universe\,\cite{Planck2015Cosm}, however its nature still remains as one of the biggest mysteries of particle physics. Even if we postulate that the dark matter is an elementary particle, its mass is allowed to be in between $10^{-22}$\,eV\,\cite{Hu:2000ke} and the Planck scale. Historically, the dark matter mass around the electroweak scale has been intensively searched for because the weakly interacting massive particle (WIMP) hypothesis has an excellent motivation. Among several strategies to search for the dark matter, the direct dark matter detection is known to be an efficient one based on the interaction between the dark matter and a nucleon, and indeed the dark matter in the well-motivated WIMP mass region\footnote{To be more precise, the WIMP mass is constrained to be in between ${\cal O}(1)$\,MeV\,\cite{Boehm:2013jpa} to ${\cal O}(100)$\,TeV\,\cite{Griest:1989wd} from the phenomenological viewpoint, though the most well-motivated mass region is ${\cal O}(10)$\,GeV to ${\cal O}(1)$\,TeV from the viewpoint of new physics models concerning the origin of the electroweak symmetry breaking.} is efficiently searched for at direct detection experiments. Since no robust signal of the dark matter is yet obtained, the experiments currently put a very severe constraint in this mass region\,\cite{Aprile:2017iyp}. As a result, the dark matter with its mass outside the well-motivated WIMP region attracts more attention than before. In particular, a light dark matter whose mass is less than a few GeV is recently being studied well, and several ideas of the direct detection to search for such a light dark matter have been proposed. For instance, the NEWS collaboration has proposed the use of various gas detectors (e.g. Helium to Xenon), which can search for a light dark matter with the mass around ${\cal O}(0.1)$\,GeV\,\cite{Profumo:2015oya}. There are even ideas to search for a very light dark matter whose mass can be as light as ${\cal O}(1)$\,keV\,\cite{Battaglieri:2017aum}.

In this paper, we propose an idea of the direct detection to search for a even lighter dark matter based on the interaction between the dark matter and a nucleon, where the dark matter mass can be much lighter than ${\cal O}(1)$\,eV. Such an ultralight dark matter is motivated because, e.g. there is a possibility to solve the so-called small scale structure crisis in our universe when its mass is as small as $10^{-22}$\,eV\,\cite{Hu:2000ke, Hui:2016ltb}. The other motivation is from its stability and abundance at the present universe. When the dynamics of the ultralight dark matter is governed by interactions suppressed by a new physics scale ($\Lambda$) at very high energy, its lifetime becomes much longer than the age of the universe. Moreover, its abundance is naturally explained by the coherent oscillation of the dark matter field\,\cite{Hui:2016ltb}: The field starts oscillating when the Hubble parameter becomes comparable to the dark matter mass ($m$) with the initial amplitude being around $\Lambda$, and the abundance is estimated to be $\Omega_{\rm DM} h^2 \sim 0.1\,(\Lambda/10^{15}\,{\rm GeV})^2\,(m/10^{-14}\,{\rm eV})^{1/2}$. Hence, we discuss the direct detection of the ultralight dark matter in the wide range of the dark matter mass and the new physics scale.

The basic idea is the use of celestial bodies (Sun, Saturn, Earth, Moon) as target martials of the direct detection of the ultralight dark matter. The bodies feel the dark matter wind because our solar system moves with respect to the rest frame of the dark matter halo, so that the scattering off the dark matter acts as a resistant force opposing their motions. In fact, there are three advantages to consider this system to detect the dark matter. First, though the momentum transfer at each scattering is tiny, ${\cal O}(m v)$ with $v \sim 250$\,km/s being the typical velocity of the dark matter around the solar system, the dark matter flux is inversely proportional to the dark matter mass $m$. The momentum transfer per unit time hence does not depend on $m$. Next, since the de Broglie wavelength of the dark matter is ${\cal O}(2\pi/(m v))$, which can be as large as the size of the celestial bodies, huge coherent enhancement is expected at the spin-independent scattering, as is usually considered in the direct WIMP detection\, \cite{Goodman:1984dc}. Finally, the resistance force affects the orbital motions of the celestial bodies, and those are precisely measured because of the astronomical ephemeris\,\cite{NASADE, IAAEPM}.

In the next section (section\,\ref{sec: scattering}), we estimate the scattering cross section between the ultralight dark matter and a celestial body. Here, one might expect the so-called ${\cal N}^2$ enhancement of the cross section with ${\cal N}$ being the number of nucleons inside the target. It is however not always true, because a simple perturbative estimate is broken down when the interaction between the dark matter and a nucleon is not very suppressed and/or the target involves a huge number of nucleons like a celestial body. We show how the non-perturbative effect can be taken into account. In section\,\ref{sec: DDofULDM}, we apply the result of the previous section to the direct detection of the ultralight dark matter. We discuss how the interaction between the dark matter and a nucleon is efficiently searched for because of the astronomical ephemeris. Finally, section\,\ref{sec: summary} is devoted to summary and discussion of our study.

\section{Scattering off a celestial body}
\label{sec: scattering}

We estimate the scattering cross section between the ultralight dark matter and a celestial body in this section. Since we are focusing on the direct detection of the dark matter, we postulate the following interaction between the dark matter ($\phi$) and a nucleon ($N$):
\begin{eqnarray}
	{\cal L}_{\rm int} = \frac{c_N}{2} \phi^2 \bar{N} N,
	\label{eq: nucleon int}
\end{eqnarray}
where the dark matter is assumed to be a boson (real scalar), because a fermionic dark matter cannot be an ultralight dark matter due to the Tremaine-Gunn bound\,\cite{Tremaine:1979we}. We impose a $Z_2$ symmetry not to have a less-dimensional interaction, $\phi \bar{N} N$, because it induces a long-range force between a pair of nucleons and already severely constrained\,\cite{Fischbach:1992fa}. The strength of the coupling constant $c_N$ depends on a concrete model behind the interaction. For instance, if the model has an interaction $\sum_q (m_q/(2 \Lambda^2)) \phi^2 \bar{q} q$ ($q$ and $m_q$ are a quark field and its mass) assuming that the exchange of a heavy particle with its mass of the new physics scale $\Lambda$ induces this quark-level interaction, the coupling constant is estimated to be \mbox{\boldmath $c_N \simeq (f_N m_N/\Lambda^2)$} with $f_N \simeq 0.3$ and $m_N$ being the nucleon mass\,\cite{Kanemura:2010sh}. Note that the scattering between the ultralight dark matter and an electron can also be considered in the same manner. Its interaction is, however, expected to be $(m_e/(2 \Lambda^2)) \phi^2 \bar{e} e$  ($e$ and $m_e$ are an electron field and its mass), and it is much smaller than that with a nucleon. Hence, we do not include the contribution from the electron scattering in this paper. With the interaction in eq.\,(\ref{eq: nucleon int}), the scattering cross section between the dark matter and a nucleon is
\begin{eqnarray}
	\frac{d \sigma_N}{d |\vec{q}|^2} = \frac{1}{4 m^2 v^2} \frac{c_N^2 m_N^2}{4 \pi (m_N + m)^2}
	\simeq \frac{c_N^2}{16 \pi m^2 v^2},
\end{eqnarray}
where $m$ and $v$ are the mass of the dark matter and the relative velocity between the dark matter and a nucleon, respectively. The momentum transfer squared is denoted by $|\vec{q}|^2$ and it takes a value between $0$ and $4 m^2 v^2$. We take the non-relativistic limit to derive the above cross section. Then, the total cross section is simply obtained to be $\sigma_N \simeq c_N^2/(4 \pi)$.

Now, let us estimate the scattering cross section between the dark matter and a target involving ${\cal N}$ nucleons according to the method used in the direct WIMP detection\,\cite{Jungman:1995df}. When we denote the wave function of the target to be $|T\rangle$ with the condition of the canonical normalization of $\langle T | T \rangle = 1$, the elastic scattering cross section is calculated to be
\begin{eqnarray}
	\frac{d \sigma_T}{d |\vec{q}|^2} \simeq \frac{c_N^2 |\langle T |{\bar N} N | T \rangle|^2}{16 \pi m^2 v^2}
	= \frac{c_N^2 {\cal N}^2 |F(q)|^2}{16 \pi m^2 v^2}
	= \frac{c_N^2 {\cal N}^2}{16 \pi m^2 v^2} \left| \int d^3 r \rho(r) e^{i \vec{q} \cdot \vec{r}} \right|^2,
	\label{eq: perturbative}
\end{eqnarray}
where we assume that the target is much heavier than a nucleon, $m_T \gg m_N$. The so-called form factor is denoted by $F(q)$ and given by the Fourier transform of the distribution function of nucleons inside the target, $\rho(r)$, where it is normalized to be $\int d^3 r \rho(r) = 1$. One might expect that the huge number of \mbox{\boldmath ${\cal N} \simeq m_T/m_N$} significantly increases the cross section, but it is not always true. {\bf The above formula cannot be applied} to the target with an extremely huge ${\cal N}$, because it is based on a perturbative calculation. Note that the critical number ${\cal N}$ leading to the breakdown depends on the strength of the coupling constant $c_N$.

Hence, we have to go beyond the perturbative calculation in order to resolve this problem. Fortunately, since we are discussing the non-relativistic scattering of a two-body system, it is generally described by an appropriate Schr\"odinger equation. Then, we obtain the correct scattering cross section between the ultralight dark matter and the target with an extremely huge ${\cal N}$ by solving the equation without relying on the perturbative calculation. With $V(r)$ being the potential generated by the target, the Schr\"odinger equation is
\begin{eqnarray}
	\left[ -\frac{\nabla^2}{2 m} + V(r) \right] \psi(\vec{r}) = E \psi(\vec{r}),
	\label{eq: schrodinger}
\end{eqnarray}
where we assume that the target is much heavier than the dark matter. The wave function of the ultralight dark matter is denoted by $\psi(\vec{r})$, and it involves information of incoming and outgoing (scattering) components. The kinetic energy carried by the incoming dark matter particle is given by \mbox{\boldmath $E = m v^2/2$}. The explicit form of the potential is obtained by comparing the cross section in eq.\,(\ref{eq: perturbative}) with the solution of the Schr\"odinger equation at the Born approximation\,\cite{SakuraiQM, MessiahQM}, which is given by $d\sigma_T/d|\vec{q}|^2 = (4 \pi v^2)^{-1} |\int d^3r V(r) e^{i \vec{q} \cdot \vec{r}} |^2$. The explicit form of the potential in the Schr\"odinger equation is hence obtained as
\begin{eqnarray}
	V(r) = - \frac{c_N}{2 m} {\cal N} \rho (r).
	\label{eq: potential}
\end{eqnarray}
The overall sign is obtained by comparing the amplitudes instead of the cross sectoins.

The intuitive understanding of the coherent enhancement (macroscopic coherence) can be obtained in the above picture using the Schr\"odinger equation. Since the target is composed of ${\cal N}$ nucleons, the potential is also rewritten by the form $V(r) = \sum_i V_i (\vec{r} - \vec{r}_i)$, where $\vec{r}_i$ is the location of the `i-th' nucleon inside the target and $V_i(\vec{r} - \vec{r}_i) = -c_N/(2 m) \delta(\vec{r} - \vec{r}_i)$.\footnote{Since nucleons inside the target is assumed to be spherically distributed, $V(r)$ depends only on $r = |\vec{r}|$.} The scattering cross section at the Born approximation is then obtained as follows:
\begin{eqnarray}
	\frac{d \sigma_T}{d |\vec{q}|^2}
	= \frac{1}{4 \pi v^2} \left| \int d^3r \sum_i V_i(\vec{r} - \vec{r}_i) e^{i \vec{q} \cdot \vec{r}_i} \right|^2
	= \frac{c_N^2}{16 \pi m^2 v^2} \left| \int d^3r {\cal N} \rho(r) e^{i \vec{q} \cdot \vec{r}} \right|^2,
\end{eqnarray}
where $\rho(r) \equiv {\cal N}^{-1} \sum_i \delta(\vec{r} - \vec{r}_i)$ is the normalized distribution function of nucleons inside the target. The coherent enhancement of the scattering cross section can be obviously seen in the above equation. When $|\vec{q} \cdot \vec{r}_i| \ll 1$ holds for any $i$, the each nucleon potential $V_i(\vec{r} - \vec{r}_i)$ additively contributes to the potential $V(r)$ and it gives the enhancement factor of ${\cal N}^2$. The additive contribution is the typical property of the spin-independent scattering. On the contrary, such an additive property is not expected at the spin-dependent scattering, because each nucleon potential depends on the nucleon spin and nucleons inside the target are usually not aligned. There is hence a cancellation among the contributions and no coherent enhancement of ${\cal N}^2$ is obtained. This picture also arrows us to intuitively understand the breakdown of the Born approximation when the target involves a huge amount of nucleons; more nucleons make the potential deeper when each contributes additively.

We are now at the position to estimate the scattering cross section between the ultralight dark matter and a celestial body, which is achieved by solving the Schr\"odinger equation beyond the Born approximation. In order to make our analysis concreate and easy, we adopt the homogeneous distribution function of nucleons inside the celestial body:
\begin{eqnarray}
	\rho(r) = \left( \frac{3}{4 \pi r_T^3} \right) \theta(r_T - r),
\end{eqnarray}
where $r_T$ is the radius of the celestial body and $\theta(x)$ is the Heaviside step function. It is then convenient to define the following dimensionless variables: $\vec{x} = m v \vec{r}$, $x_T \equiv m v r_T$ and $y \equiv [3 |c_N| {\cal N}/(4 \pi r_T)]^{1/2}$, where $x_T$ is the de Broglie wavelength with respect to the size of the celestial body $r_T$, while $y$ represents the strength of the interaction between the ultralight dark matter and the celestial body, or in other words, the depth of the potential created by the body. With the variables, the Schr\"odinger equation is rewritten as
\begin{eqnarray}
	-\left[ \nabla_x^2 + {\rm sign}(c_N) \frac{y^2}{x_T^2} \theta(x_T - |\vec{x}|) \right] \psi(\vec{x}) = \psi(\vec{x}).
	\label{eq: dimension-less schrodinger}
\end{eqnarray}

\begin{figure}[t]
	\centering
	\includegraphics[width=0.47\textwidth]{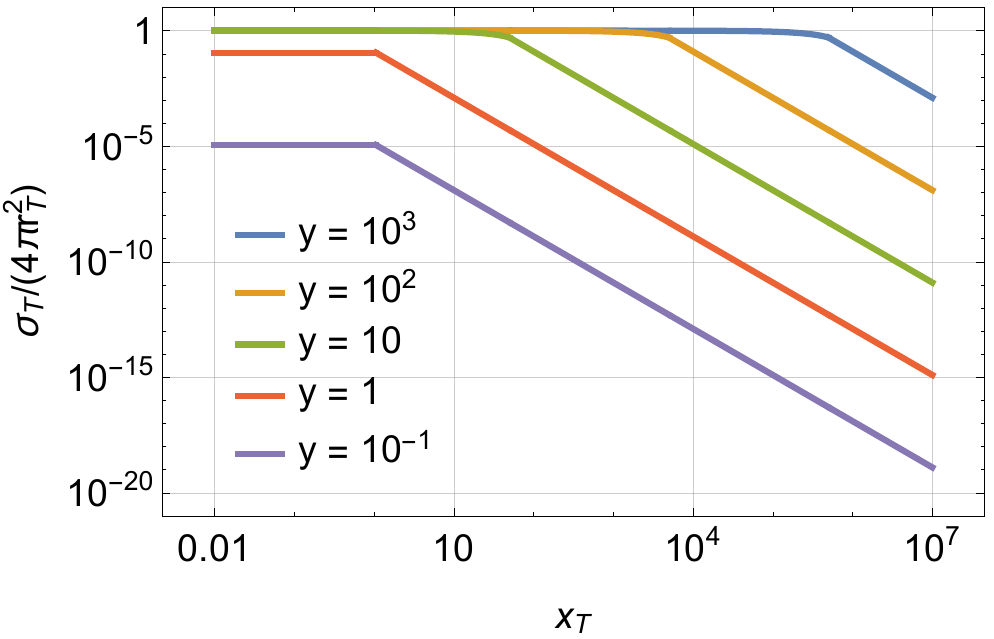}
	\qquad
	\includegraphics[width=0.47\textwidth]{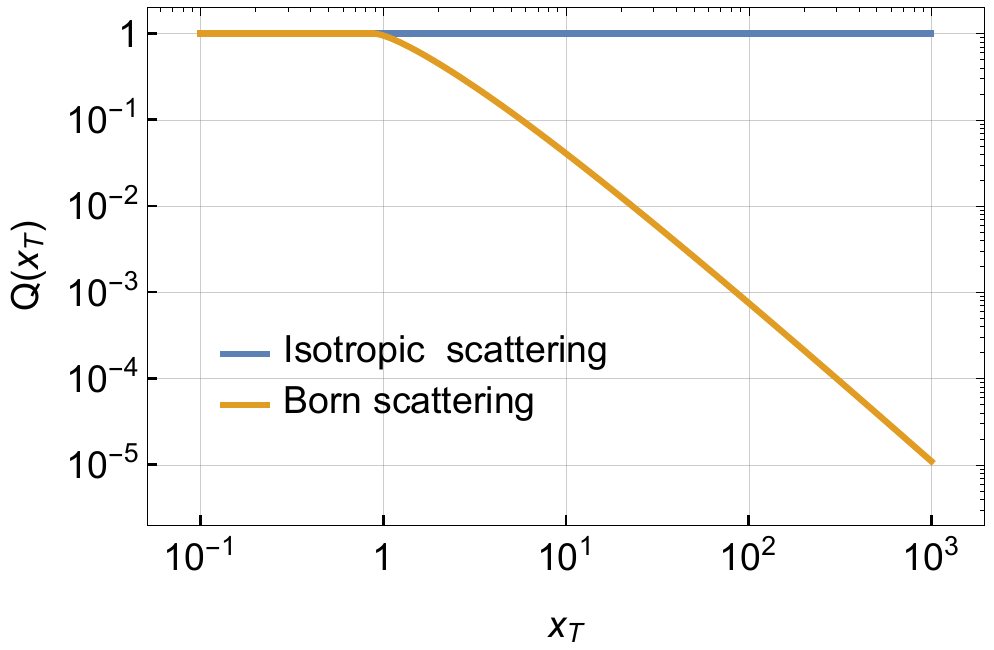}
	\caption{\sl \small{\bf Left panel:} The total scattering cross section (normalized by $4 \pi r_T^2$) between the ultralight dark matter and a celestial body as a function of $x_T (= m v r_T)$ with $y (= [3 |c_N| {\cal N}/(4 \pi r_T)]^{1/2})$ being fixed as $0.1$, $1$, $10$, $10^2$ or $10^3$. {\bf Right panel:} The efficiencies of the momentum transfer in eq.\,(\ref{eq: momentum transfer}) for the cases of isotropic and Born scatterings as a function of $x_T$. See text for more details.}
	\label{fig: Total_Efficiency}
\end{figure}

Let us first estimate the total scattering cross section $\sigma_T$. As we have mentioned above, the scattering cross section is well described by the Born approximation when the interaction is weak enough (in other words, $y$ is small enough). Its explicit form is obtained by integrating the differential scattering cross section\,(\ref{eq: perturbative}) over the momentum transfer $|\vec{q}|^2$:
\begin{eqnarray}
	\sigma_T^{(B)}(y, x_T) \simeq 4 \pi r_T^2 y^4
	\left[
		\frac{1}{9} \theta \left( \sqrt{\frac{9}{8}} - x_T \right) + \frac{1}{8 x_T^2} \theta \left( x_T - \sqrt{\frac{9}{8}} \right)
	\right],
\end{eqnarray}
where the Born approximation is validated in the domain of $y \lesssim \sqrt{2}$\,\cite{LandauLifshitz198101}. On the other hand, when the interaction is not weak, the Born approximation can be applied only in the large $x_T$ domain, namely $x_T \gtrsim y^2/2$\,\cite{LandauLifshitz198101}. On the contrary, in the small $x_T$ domain $x_T \lesssim 1$, $\sigma_T$ is dominated by the s-wave contribution, and its explicit form is estimated to be
\begin{eqnarray}
	\sigma_T^{(S)}(y) \simeq 4 \pi r_T^2
	\left[ \frac{y^4}{9} \theta \left( 3 - y^2 \right) + \theta \left( y^2 - 3\right) \right].
\end{eqnarray}
In the middle $x_T$ domain, though it is difficult to compute the total scattering cross section analytically, there are some naive estimates. For instance, in the domain of $y/\sqrt{2} \lesssim x_T \lesssim y^2/2$, the so-called Eikonal approximation is used to estimate the cross section, and it leads to $\sigma_T \sim 2 \pi r_T^2$\,\cite{SakuraiQM}. In the other middle $x_T$ domain, $1 \lesssim x_T \lesssim y/\sqrt{2}$, the cross section is estimated to be in between $2 \pi r_T^2$ and $4 \pi r_T^2$ assuming that the depth of the potential $y$ is much larger than $x_T$\,\cite{doi:10.1063/1.1722537}. In any case, whenever $x_T$ is smaller than $y^2/2$, the cross section turns out to be of the order of $\pi r_T^2$, so that we estimate $\sigma_T$ using the following formula:
\begin{eqnarray}
	\sigma_T &\simeq& \sigma_T^{(S)}\cdot\theta (1 - x_T) + \sigma_T^{(SB)}\cdot\theta (x_T - 1) \theta (x_T - y^2/2)
	+ \sigma_T^{(B)}\cdot\theta (x_T - \max [1, y^2/2]),
	\nonumber \\
	\sigma_T^{SB} &\equiv&
	[ \{ \sigma_T^{(B)}(y, y^2/2) - \sigma_T^{(S)}(y) \}/(y^2/2 - 1) ] (x - 1) + \sigma_T^{(S)}(y),
\end{eqnarray}
where it can be used without respect to the overall sign of the potential in eq.\,(\ref{eq: dimension-less schrodinger}). The total cross section $\sigma_T$ normalized by $4 \pi r_T^2$ is depicted in Fig.\,\ref{fig: Total_Efficiency} (left panel) as a function of $x_T$ with $y$ being fixed as $0.1$, $1$, $10$, $10^2$ or $10^3$. The figure shows that the cross section is saturated by $4 \pi r_T^2$ when the depth of the potential is large enough ($y \gg 1$) unlike a naive expectation about the coherent enchantment of $\sigma_T$ seen at the Born approximation.

Next, let us discuss the $|\vec{q}|^2$ dependence of the cross section. The most important physical quantity here is the efficiency of the momentum transfer rather than the differential scattering cross section itself, which appears in the formula of a resistant force opposing motion of the celestial body, as shown in the next section. The efficiency is defined as
\begin{eqnarray}
	Q(x_T) \equiv \frac{1}{2 x_T^2} \int_0^{4 x_T^2} dz^2 z^2
	\left[ \frac{1}{\sigma_T} \frac{d \sigma_T}{d z^2} \right],
	\label{eq: momentum transfer}
\end{eqnarray}
where $z^2 \equiv |\vec{q}|^2 r_T^2$. When the scattering proceeds isotropically such as the s-wave contribution, the efficiency gives one without depending on $x_T$. On the other hand, when the differential scattering cross section favors the forward scattering like the case of the Born approximation in eq.\,(\ref{eq: perturbative}), the efficiency gives a value smaller than one. The efficiencies for these two cases (isotropic and Born scatterings) are shown in Fig.\,\ref{fig: Total_Efficiency} (right panel) as a function of $x_T$. In the middle $x_T$ domain with $y \gg 1$, the naive estimates mentioned above suggests that the efficiency is in between those of isotropic and Born scatterings. We apply $Q(x_T)$ of the Born scattering also to this domain, because it is difficult to compute $Q(x_T)$ correctly in this domain. It gives a smaller efficiency than the actual one, making the constraint on the interaction between the ultralight dark matter and a celestial body (discussed in the next section) conservative. The explicit form of $Q(x_T)$ for the Born scattering is
\begin{eqnarray}
	Q^{(B)}(x_T) \simeq \theta(3 - 4 x_T^2)
	+ \frac{3}{8 x_T^2 - 3} \left[ 1 + 2 \ln \left( \frac{4 x_T^2}{3} \right) \right] \theta(4 x_T^2 - 3).
\end{eqnarray}
Since $Q^{(B)}(x_T)$ becomes one in the small $x_T$ domain as seen in Fig.\,\ref{fig: Total_Efficiency} (right panel), it can also be applied to this $x_T$ domain. After all, we can use it in the entire range of $x_T$ and $y$.

\section{Direct detection of Ultralight dark matter}
\label{sec: DDofULDM}

The physical quantity that we focus on in this section is the deceleration of a celestial body caused by a resistant force opposing motion due to its scattering off the ultralight dark matter. We first discuss how the deceleration parameter is described by the scattering cross section $\sigma_T(y, x_T)$ and the efficiency of the momentum transfer $Q(x_T)$ developed in the previous section. We next summarize observational limits on the deceleration parameter obtained from the astronomical ephemeris. Finally, with the use of the results in these two subsections, we put a limit on the spin-independent scattering cross section $\sigma_N$.

\subsection{Deceleration of a celestial body}
\label{subsec: deceleration parameter}

Taking the rest frame of a celestial body and assuming that the mass of the body ($m_T$) is much larger than the ultralight dark matter mass ($m$), the deceleration vector is
\begin{eqnarray}
	{\vec a}_0 = \frac{\rho_{\rm DM}}{m\,m_T} \int d^3 v f(\vec{v})
	\int d|\vec{q}|^2 \, \vec{q} \, \frac{d\sigma}{d|\vec{q}|^2} v
	=
	\vec{n}_z \frac{\rho_{\rm DM}}{m_T} \int d^3v f(\vec{v}) \, v_z \, v \, \sigma_T(y, x_T) \, Q(x_T),
	\label{eq: deceleration vector}
\end{eqnarray}
where $\vec{v}$ is the velocity of the dark matter particle and $v = |\vec{v}|$. We assume that the dark matter wind is mainly from the (positive) $z$ direction, so that $\vec{n}_z$ stands for a unit vector along the $z$ direction. The mass density of the dark matter particle at the solar system is denoted by $\rho_{\rm DM} \simeq 0.3$\,GeV $c^{-2}$\,cm$^{-3}$, while its velocity distribution is described by $f(\vec{v})$ with the normalization being $\int d^3 v\,f(\vec{v}) = 1$. The explicit form of $f(\vec{v})$ is given by
\begin{eqnarray}
	f(\vec{v}) \simeq \left( \frac{3}{2 \pi v_0^2} \right)^{3/2}
	\exp \left[ - \frac{3}{2 v_0^2} \left\{ v_x^2 + v_y^2 + (v_z + v_T)^2 \right\} \right]
	\theta \left( v_{\rm esc}^2 - \left\{ v_x^2 + v_y^2 + (v_z + v_T)^2 \right\} \right).
	\label{eq: velocity distribution function}
\end{eqnarray}
We adopt the standard halo model for $f(\vec{v})$\,\cite{Lewin:1995rx} neglecting the proper motion of each celestial body for simplicity, where the velocity of the solar system with respect to the rest frame of the dark matter halo, the velocity dispersion and the escape velocity of the dark matter particle in the halo are $v_T = 244$\,km/s, $v_{\rm esc} = 600$\,km/s and $v_0 \simeq 230$\,km/s, respectively.

\begin{figure}[t]
	\centering
	\includegraphics[width=0.47\textwidth]{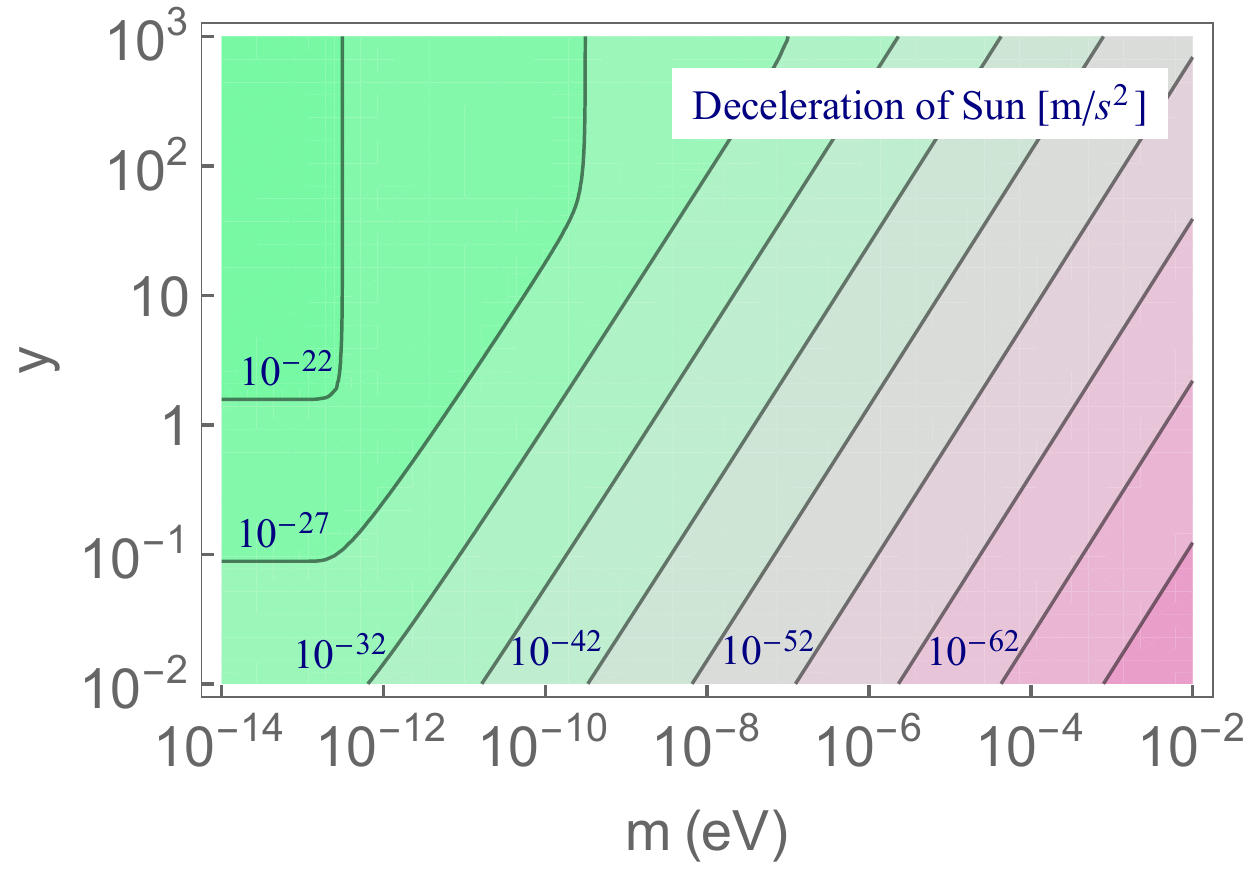}
	\qquad
	\includegraphics[width=0.47\textwidth]{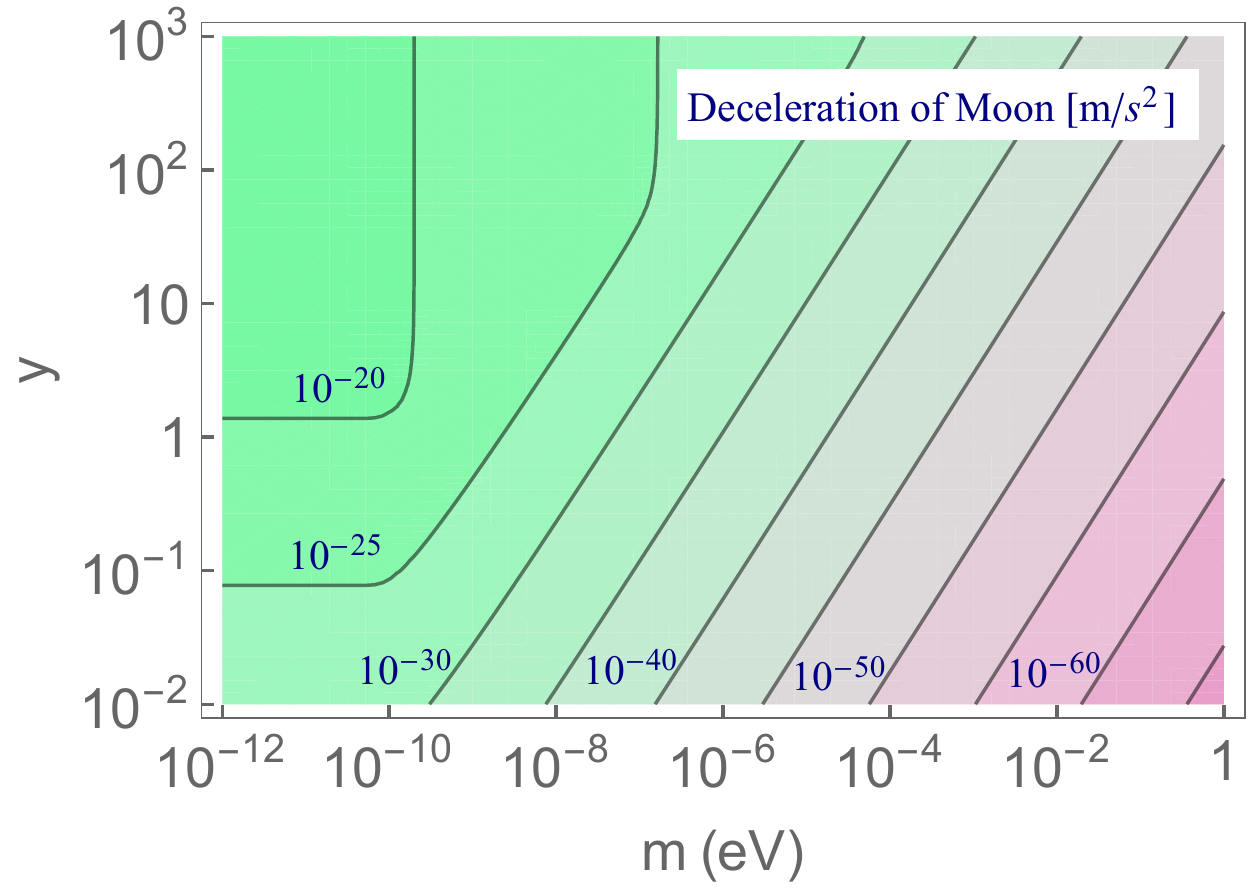}
	\caption{\sl \small The deceleration parameter $a_0$ for the sun (left panel) and the moon (right panel), caused by their scatterings off ultralight dark matter particles in the halo, on the plane of $(m, y)$.}
	\label{fig: deceleration parameter}
\end{figure}

The deceleration parameter \mbox{\boldmath $a_0 \equiv |\vec{a}_0|$} is given by the integration of the product of the two components, $f(\vec{v})\,v_z\,v$ and $\sigma_T(y, x_T)\,Q(x_T)$, over the velocity of the dark matter particle $v$. Because the former factor has a non-negligible value only at around $v \sim {\cal O}(100)$\,km/s while the latter one is constant at $x_T = 2 m v r_T \lesssim 1$, the deceleration parameter $a_0$ becomes independent of the dark matter mass when $m \lesssim 10^{-10}\,(10^3\,{\rm km}/r_T)$\,eV. The deceleration parameter in this mass region is analytically expressed by the following formula:
\begin{eqnarray}
	a_0 &\simeq& \frac{\rho_{\rm DM}}{m_T}\,v_T\,v_0\,\sigma_T^{(S)}(y)\,g\left( \frac{v_T}{v_0} \right),
	\nonumber \\
	g(x)
	&=&
	\frac{1}{9 x^3}
	\left[ \sqrt{ \frac{6}{\pi} } x ( 1 + 3 x^2) \exp \left( -\frac{3 x^2}{2} \right) 
	- \left( 1 - 6 x^2 - 9 x^4 \right) {\rm Erf} \left( \sqrt{ \frac{3}{2} } x \right) \right],
	\label{eq: asymptotic deceleration parameter}
\end{eqnarray}
The deceleration parameter $a_0$ for the sun and the moon caused by their scatterings off ultralight dark matter particles in the halo are shown in the left and the right panels of Fig.\,\ref{fig: deceleration parameter}, respectively, on the plane of $(m, y)$. As expected, the parameter $a_0$ becomes independent of $m$ when it is small enough, and is described very well by the formula in eq.\,(\ref{eq: asymptotic deceleration parameter}).

This is, however, not the end of the story concerning the deceleration parameter. When the ultralight dark matter is very smoothly distributed at the scale much smaller than that of the celestial bodies according to the velocity distribution in eq.\,(\ref{eq: velocity distribution function}), a further enhancement of the deceleration is expected from the so-called stimulate emission effect of the ultralight dark matter particle. The deceleration vector $\vec{a}_0$ in eq.\,(\ref{eq: deceleration vector}) is then modified as
\begin{eqnarray}
	{\vec a} = \frac{\rho_{\rm DM}}{m\,m_T} \int d^3 v f(\vec{v})
	\int d|\vec{q}|^2 \, \vec{q} \, \frac{d\sigma}{d|\vec{q}|^2} v
	\left[ 1 + \rho_{\rm DM} \frac{(2\pi)^3}{m^4} f(\vec{v}') \right],
	\label{eq: deceleration vector with s}
\end{eqnarray}
where $\vec{v}'$ is the velocity of the dark matter particle after the scattering. Since the enhancement factor is estimated to be $(\rho_{\rm DM}/m)\,(2\pi)^3/(m\,v_0)^3 \sim 10^{37} (10^{-10}\,{\rm eV}/m)^4$, the deceleration parameter is significantly boosted when the dark matter is much lighter than ${\cal O}(1)$\,eV. It is, however, worth notifying that the boost is suppressed when the dark matter is not smoothly distributed at the small scale\,\cite{Guth:2014hsa, Chakrabarty:2017fkd}. In such a case, it becomes difficult to find dark matter particles which have the same quantum number (velocity) as the one that the dark matter particle scattered off has. As a result, the deceleration parameter $a_0$ estimated in eq.\,(\ref{eq: deceleration vector with s}) should be regarded as the maximal deceleration that we can expect, while that in eq.\,(\ref{eq: deceleration vector}) gives the minimal one. Since the distribution of the ultralight dark matter at the small scale is an issue still under debate, we simply introduce the so-called boost factor $B$ in order to take this effect into account in our analysis. The factor is then defined as
\begin{eqnarray}
	|\vec{a}| = B\,|\vec{a}_0|,
\end{eqnarray}
where the boost factor takes a value between $1 \leq B \lesssim (\rho_{\rm DM}/m)\,(2\pi)^3/(m\,v_0)^3$ at each mass of the ultralight dark matter. Since the parameter $y$ is related to the scattering cross section between the dark matter and a nucleon through the equation $y = [3 |c_N| {\cal N}/(4 \pi r_T)]^{1/2} = [9 {\cal N}^2/(4 \pi r_T^2)]^{1/4} \sigma_N^{1/4}$ with ${\cal N} \simeq m_T/m_N$, the constraint on the deceleration parameter $|\vec{a}|$ obtained from the astronomical ephemeris is translated into that on the scattering cross section $\sigma_N$. We will see in the following sections that the precise astronomical ephemeris already put a severe constraint on $\sigma_N$ even if the boost factor $B$ is not large.

\subsection{Limits from the astronomical ephemeris}
\label{subsec: astronomical ephemeris}

The DE from NASA, USA\,\cite{NASADE} and the EPM from IAA RAS, Russia\,\cite{IAAEPM} are known to be high precision ephemerides of solar system bodies, where the theoretical prediction about the motion of the bodies are compared with a huge amount of observational data numerically. We should hence implement the deceleration effect of the bodies (due to the ultralight dark matter) into these ephemeris codes to precisely evaluate the effect, though it is beyond the scope of this paper. Instead, we simply estimate the effect based on the following discussion. First, since the motion of the solar system with respect to the rest frame of the dark matter halo is much faster than the proper motion of each celestial body\,\cite{Lewin:1995rx}, we only consider the deceleration effect caused by the motion of the solar system. Next, focusing on the gravitationally bounded two-body system such as the sun and the earth, or the earth and the moon, we estimate the deceleration effect assuming that the dark matter wind comes from the direction parallel to the orbital plane of the system.\footnote{Since the orbital plain is usually tilted by some degrees with respect to the direction of the solar system's move\,\cite{Lewin:1995rx}, the resistance force towards the direction parallel to the plane becomes weaker. We, however, neglect such a effect, for it does not change the result of our order estimate. The resistance force towards the direction perpendicular to the orbital plane leads to a harmonic oscillation of the orbit, and it does not gives a sizable effect on the modification of the orbit compared to the force towards the parallel direction.} Since the resistant force caused by the dark matter is much smaller than the central force binding celestial bodies in the system, the deceleration effect modifies the orbital motion perturbatively. To good approximation, the celestial body is in the circular motion when the deceleration effect does not exists, so that the velocity of the body ($v_*$) and the radius of the circular motion ($l_*$) are individually preserved. When the deceleration effect exists, the velocity is gradually modified as $dv_*/dt \sim |\vec{a}|$. The celestial body is, however, still gravitationally bounded in the system, so that it follows the Kepler's low ($l_*\,v_* \sim$ const.) at certain extent in the perturbative treatment of the resistance force. As a result, the radius of the motion is modified as
\begin{eqnarray}
	dl_* \sim dv_* \times (l_*/v_*) \sim a\,T\times (l_*/v_*),
	\label{eq: constraint formula}
\end{eqnarray}
where $T$ is the duration of the observation at the astronomical ephemerides. Note here that $dl_*$ does not means the average of the modification of the orbital motion from the circular one but the maximal one caused by the resistance force during the observation.

The astronomical ephemerides addressed above put severe constraints on $dl_*$, the modification of the orbital motion. Since dynamics of the solar system observed so far is consistent with known physics, we use the uncertainty of the distance between celestial bodies in the bounded system as an upper limit on $dl_*$. According to references\,\cite{NASADE, Krasinsky2004, standish_2004}, the limit on $dl_*$ for each bounded system as well as its $v_*$, $l_*$ and $T$ are summarized as follows:
\vspace{0.7cm}
\begin{table}[h]
	\begin{center}
		\begin{tabular}{r|lccc}
			& $dl_*$\,[$m$] & $v_*$\,[$m/s$] & $l_*\,[$m$] $ & $T\,[s]$ \\
			\hline
			Earth-Sun & $1 \times 10$ & $3 \times 10^4$ & $2 \times 10^{11}$ & $3 \times 10^9$ \\
			Saturn-Sun & $5 \times 10$ & $1 \times 10^4$ & $1 \times 10^{12}$ & $3 \times 10^8$ \\
			Moon-Earth & $8 \times 10^{-3}$ & $1 \times 10^3$ & $4 \times 10^8$ & $1 \times 10^9$ \\
			\hline
		\end{tabular}
	\end{center}
\end{table}

\noindent
These upper limits on the modification $dl_*$ are directly translated into those on the deceleration parameter $|\vec{a}|$ through the estimate in eq.\,(\ref{eq: constraint formula}). We will see in the next subsection that how strongly these limits put a severe constraint on the strength of the interaction (namely, the scattering cross section) between the ultra-light dark matter and a nucleon.

\subsection{Constraint on $\sigma_N$}

\begin{figure}[t]
	\centering
	\includegraphics[width=0.47\textwidth]{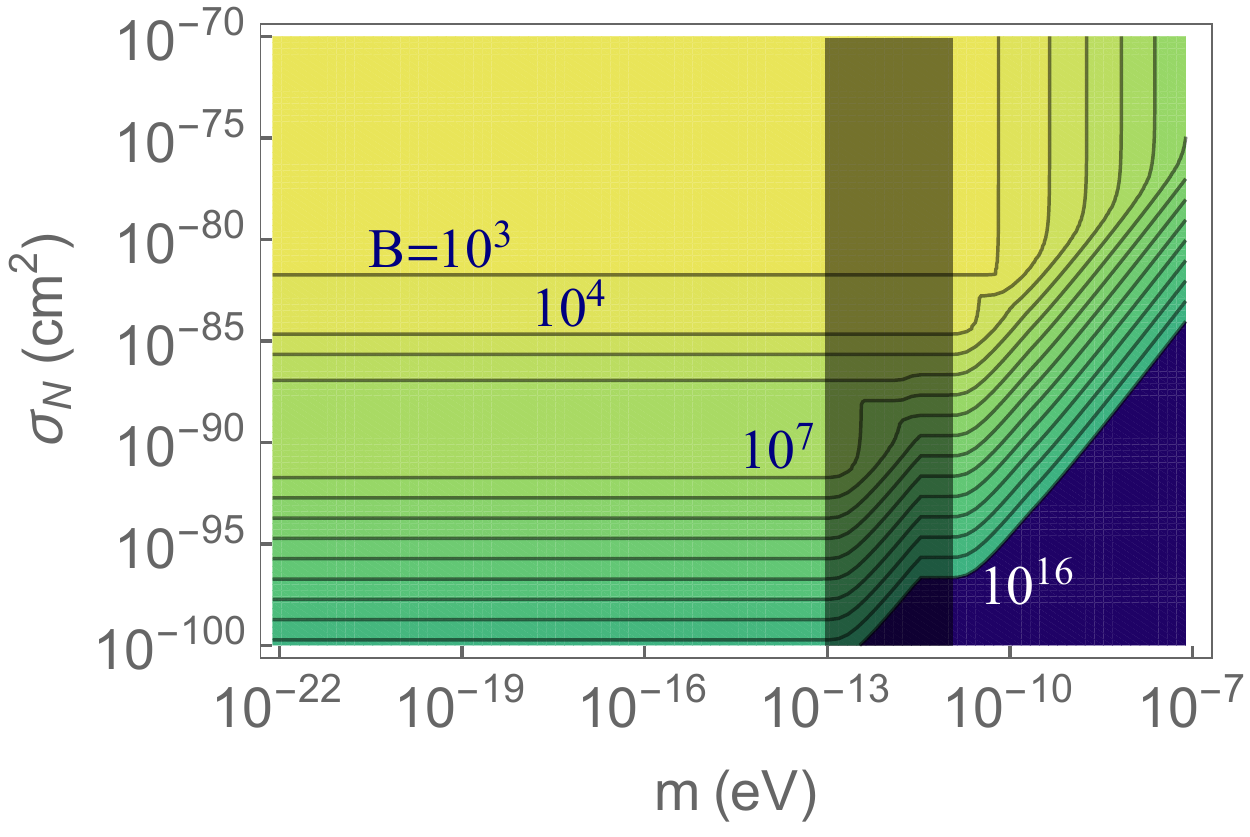}
	\qquad
	\includegraphics[width=0.47\textwidth]{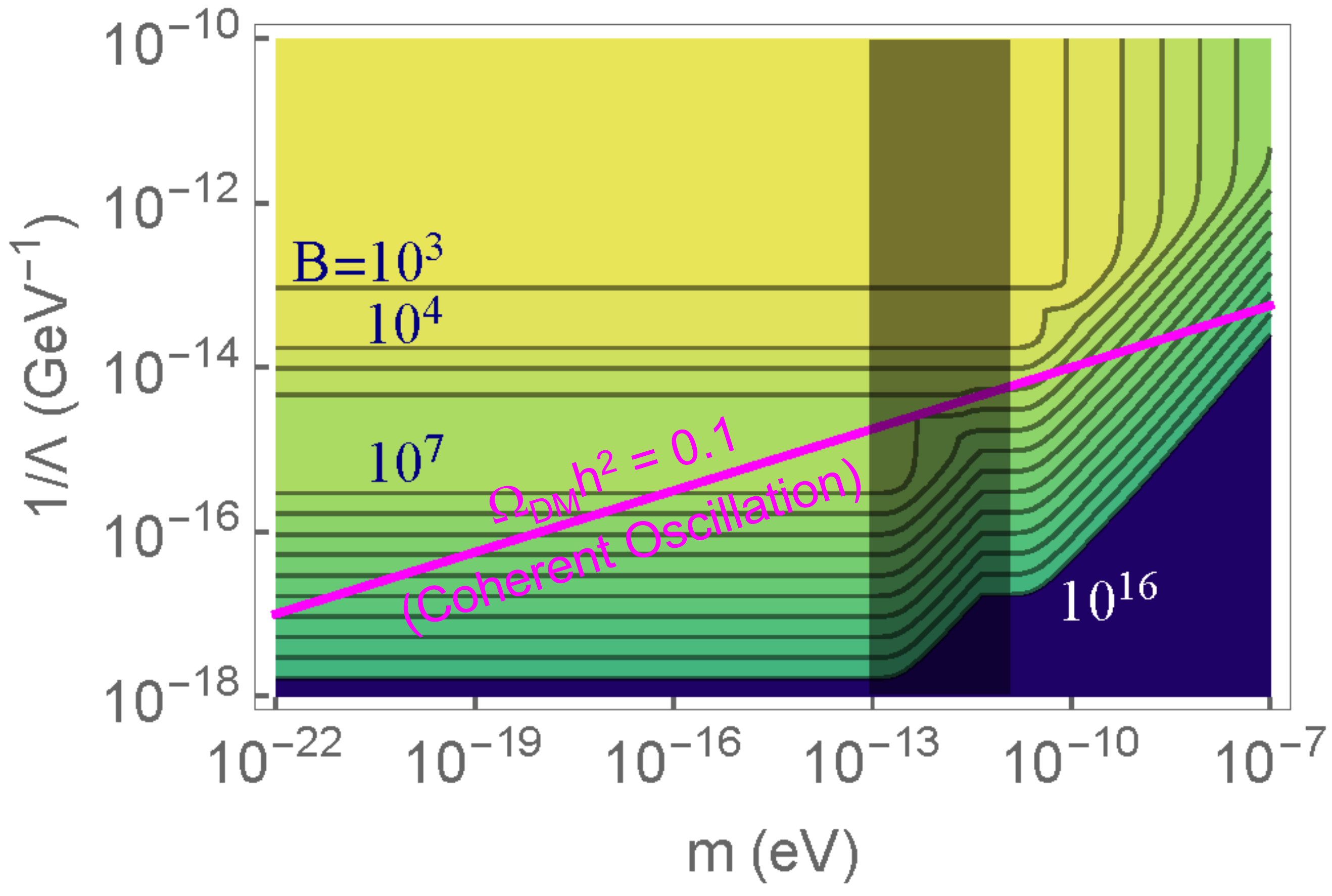}
	\caption{\sl \small {\bf Left panel:} Constraint on the scattering cross section between the ultralight dark matter and a nucleon ($\sigma_N$) as a function of the ultralight dark matter mass ($m$). Solid lines are constraints obtained with the boost factor $B$ being $10^3$ to $10^{16}$ by one order of magnitude. {\bf Right panel:} The same plot as the left one using the new physics scale $\Lambda$ instead of $\sigma_N$, where these two variables are related through the equation, $\sigma_N = c_N^2/(4\pi)$ with $c_N = f_N m_N/\Lambda^2$, mentioned in Sec.\,\ref{sec: scattering}. See text for details of the black-shaded region in both panels as well as the magenta line in the right panel.}
	\label{fig: constraints}
\end{figure}

We are now at the position to put a constraint on $\sigma_N$, the scattering cross section between the ultralight dark matter and a nucleon. By comparing the deceleration parameter $|\vec{a}|$ estimated in Sec.\,\ref{subsec: deceleration parameter} and the limits from the astronomical ephemeris in Sec.\,\ref{subsec: astronomical ephemeris}, we obtain the constraint on the cross section $\sigma_N$ as shown in the left panel of Fig.\,\ref{fig: constraints}, where solid lines are constraints as a function of the dark matter mass obtained with the boost factor being $10^3$ to $10^{16}$ by one order of magnitude. The mass region $10^{-13}$\,eV $\lesssim m \lesssim$ $10^{-11}$\,eV in the figure is shaded by black, as it is already excluded by the so-called back hole superradiance\,\cite{Arvanitaki:2014wva, Cardoso:2018tly}: The superradiant instability of spinning black holes exists due to the presence of the ultralight dark matter field and it extracts angular momentum from the black holes, putting a limit on the maximum spin of astrophysical black holes.\footnote{If the mass of the dark matter is about $6 \times 10^{-13}$\,eV, it may be possible to explain the spin-less nature of the black holes observed at the advanced LIGO utilizing the effect of the black hole superradiance\,\cite{MukaidaYanagida}.} In order to intuitively understand which energy scale the astronomical ephemeris is exploring, we also show the same plot in the right panel of the figure with the new physics scale $\Lambda$ instead of the cross section $\sigma_N$ using the relation, $\sigma_N = c_N^2/(4\pi)$ with $c_N = f_N m_N/\Lambda^2$, addressed in the beginning of Sec.\,\ref{sec: scattering}. The magenta line in the figure is a theoretical prediction assuming that the abundance of the ultralight dark matter at the present universe is determined by the coherent oscillation of the dark matter field with the initial amplitude of $\Lambda$, as addressed in introduction.

It can be seen that the astronomical ephemeris has a strong capability to put a constraint on $\sigma_N$ even when the boost factor $B$ is as small as $10^3$. The qualitative feature of the constraint can be understood as follows: When the cross section $\sigma_N$ is large enough, the constraint comes from a smaller celestial body, for the scattering cross section between the celestial body and the dark matter is saturated by the geometrical factor and thus the deceleration parameter becomes proportional to $r_T^2/m_T \propto r_T^{-1}$ with $r_T$ and $m_T$ being the radius and the mass of the body, respectively. On the other hand, when $\sigma_N$ is not large, a larger celestial body gives a severer constraint, because the scattering cross section is boosted by the coherent enhancement and thus $|\vec{a}| \propto {\cal N}^2/m_T \propto r_T^3$. It is also interesting to see that the new physics scale can be explored up to the Planck scale, $\Lambda \sim 10^{18}$\,GeV, when the dark matter mass is less than $10^{-13}$\,eV and the boost factor is larger than $10^{16}$.

\section{Summary and Discussion}
\label{sec: summary}

We have proposed a novel idea of the direct detection to search for the ultralight dark matter based on the interaction between the dark matter and a nucleon; the precise astronomical ephemeris has a strong capability to detect such a light dark matter, as celestial (solar system) bodies feel the dark matter wind and it acts as a resistant force opposing their motions. We have estimated the resistant force (deceleration parameter) based on the calculation of the elastic scattering cross section between the dark matter and the celestial bodies beyond the Born approximation, and shown that the astronomical ephemeris indeed put a strong constraint on the interaction between the dark matter and a nucleon. The idea observing the resistance force can also be applied to other targets. For example, a torsion balance experiment on the ground will observe the force more accurately\,\cite{Hagmann:1999kf}, and it enables us to detect the ultralight dark matter with mass heavier than $10^{-18}$\,eV. Another example is a gravitational experiment (e.g. the LISA project\,\cite{Seoane:2013qna}), where the force could be measured very accurately within the certain range of the ultralight dark matter mass.

We have also pointed out that the constraint depends on how smoothly the ultralight dark matter is distributed at the scale smaller than the celestial bodies because of the stimulated emission effect. Thanks to the macroscopic coherence of the scattering between the dark matter and a celestial body, the resistance force becomes observable even if the boost factor caused by the stimulated emission effect is not large. On the other hand, this fact also means that, if the boost factor is huge, it may be even possible to observe the scattering without the macroscopic coherence like the case of axion like particles. According to the analysis developed in this paper, we obtain the following perspective. Suppose that the ultralight dark matter does not have the scalar interaction in eq.\,(\ref{eq: nucleon int}) but has some other interactions with a nucleon, the dark matter is expected to have, at least, the scattering cross section of ${\cal O}(M_{\rm pl}^{-2}) \sim 10^{-76}$\,km$^2$ with a celestial body, which is $10^{82{\rm -}87}$ times smaller than that we have discussed in this paper. On the other hand, when the dark matter is smoothly distributed at the small scale, the boost factor can be as large as $10^{37}\,(10^{-10}\,{\rm eV}/m)^4$. As a result, the resistance force becomes observable if the dark matter mass is as small as $10^{-22}$\,eV.

There are a few discussions which makes our analysis accurate. First one is the effect of the inelastic scattering between the ultralight dark matter and a celestial body. Some readers might worry about that the inelastic scattering cross section ($\sigma_r$) becomes larger than the elastic scattering cross section ($\sigma_e$) and suppresses the resistance force evaluated in this paper. Fortunately, according to a generic property of the quantum mechanics\,(Sec.\,139 in Ref.\,\cite{LandauLifshitz198101}), a situation $\sigma_r \gg \sigma_e$ occurs only when both elastic and inelastic interactions are small. Then, we can use the Born approximation and the elastic scattering is not affected by the existence of the inelastic scattering, for these two processes are not interfered with each other. On the other hand, when the inelastic scattering becomes larger, the elastic scattering also becomes larger. In other words, the presence of the inelastic scattering necessarily involves that of the elastic scattering at the same time, and the certain strength of the resistance force is expected to be obtained. Moreover, when the inelastic scattering is large, a large portion of the energy used to alter the internal structure of the celestial body will make the body heated, and it gives another severe constraint on the interaction between the dark matter and a nucleon\,\cite{Kawasaki:1991eu}. The other discussion is about the initial state of the ultralight dark matter. In our analysis, the ultralight dark matter was assumed to preserve a particle picture, and each scattering process was treated independently. On the other hand, if the initial state of the dark matter is much entangled quantum mechanically, our formula would be required to be modified. We leave this problem for future study.

\vskip 0.5cm
\noindent
{\bf Acknowledgments}\\[0.1cm]
\noindent
We would like to thank M. Ibe for fruitful discussion about the unitarity limit of the scattering cross section and H. Murayama for valuable discussion about the stimulate emission effect of a light dark matter. We would also like to thank S. Shirai for turning our eyes to a gravitational experiment as an alternative target that our study can be applied to. This research was supported by the Grant-in-Aid for Scientific Research from the Ministry of Education, Culture, Science, Sports, and Technology (MEXT), Japan No. 17H02878 (S. M. \& T. T. Y.), 16H02176 (S. M. \& T. T. Y.), 26104009 (S. M. \& T. T. Y.) and No. 26104001 (T. T. Y.), and also the World Premier International Research Center Initiative (WPI Initiative), MEXT, Japan. The study of H.F. is supported in part by the Research Fellowship for Young Scientists from the Japan Society for the Promotion of Science (JSPS).

\bibliographystyle{./utphys}
\bibliography{refs}

\providecommand{\href}[2]{#2}\begingroup\raggedright\begin{thebibliography}{10}

\bibitem{Planck2015Cosm}
{\bfseries Planck} Collaboration, P.~A.~R. Ade {\em et~al.}, ``{Planck 2015
  results. XIII. Cosmological parameters},''
  \href{http://dx.doi.org/10.1051/0004-6361/201525830}{{\em Astron. Astrophys.}
  {\bfseries 594} (2016) A13},
\href{http://arxiv.org/abs/1502.01589}{{\ttfamily arXiv:1502.01589
  [astro-ph.CO]}}.

\bibitem{Hu:2000ke}
W.~Hu, R.~Barkana, and A.~Gruzinov, ``{Cold and fuzzy dark matter},''
  \href{http://dx.doi.org/10.1103/PhysRevLett.85.1158}{{\em Phys. Rev. Lett.}
  {\bfseries 85} (2000) 1158--1161},
\href{http://arxiv.org/abs/astro-ph/0003365}{{\ttfamily arXiv:astro-ph/0003365
  [astro-ph]}}.

\bibitem{Boehm:2013jpa}
C.~Boehm, M.~J. Dolan, and C.~McCabe, ``{A Lower Bound on the Mass of Cold
  Thermal Dark Matter from Planck},''
  \href{http://dx.doi.org/10.1088/1475-7516/2013/08/041}{{\em JCAP} {\bfseries
  1308} (2013) 041},
\href{http://arxiv.org/abs/1303.6270}{{\ttfamily arXiv:1303.6270 [hep-ph]}}.

\bibitem{Griest:1989wd}
K.~Griest and M.~Kamionkowski, ``{Unitarity Limits on the Mass and Radius of
  Dark Matter Particles},''
\href{http://dx.doi.org/10.1103/PhysRevLett.64.615}{{\em Phys. Rev. Lett.}
  {\bfseries 64} (1990) 615}.

\bibitem{Aprile:2017iyp}
{\bfseries XENON} Collaboration, E.~Aprile {\em et~al.}, ``{First Dark Matter
  Search Results from the XENON1T Experiment},''
\href{http://arxiv.org/abs/1705.06655}{{\ttfamily arXiv:1705.06655
  [astro-ph.CO]}}.

\bibitem{Profumo:2015oya}
S.~Profumo, ``{GeV dark matter searches with the NEWS detector},''
  \href{http://dx.doi.org/10.1103/PhysRevD.93.055036}{{\em Phys. Rev.}
  {\bfseries D93} no.~5, (2016) 055036},
\href{http://arxiv.org/abs/1507.07531}{{\ttfamily arXiv:1507.07531 [hep-ph]}}.

\bibitem{Battaglieri:2017aum}
M.~Battaglieri {\em et~al.}, ``{US Cosmic Visions: New Ideas in Dark Matter
  2017: Community Report},''
\href{http://arxiv.org/abs/1707.04591}{{\ttfamily arXiv:1707.04591 [hep-ph]}}.

\bibitem{Hui:2016ltb}
L.~Hui, J.~P. Ostriker, S.~Tremaine, and E.~Witten, ``{Ultralight scalars as
  cosmological dark matter},''
  \href{http://dx.doi.org/10.1103/PhysRevD.95.043541}{{\em Phys. Rev.}
  {\bfseries D95} no.~4, (2017) 043541},
\href{http://arxiv.org/abs/1610.08297}{{\ttfamily arXiv:1610.08297
  [astro-ph.CO]}}.

\bibitem{Goodman:1984dc}
M.~W. Goodman and E.~Witten, ``{Detectability of Certain Dark Matter
  Candidates},''
\href{http://dx.doi.org/10.1103/PhysRevD.31.3059}{{\em Phys. Rev.} {\bfseries
  D31} (1985) 3059}.

\bibitem{NASADE}
W.~M. Folkner, J.~G. Williams, D.~H. Boggs, R.~S. Park, and P.~Kuchynka, ``The
  planetary and lunar ephemerides de430 and de431,'' {\em Interplanet. Netw.
  Prog. Rep} {\bfseries 196} (2014) 1--81.
  \url{https://ipnpr.jpl.nasa.gov/progress_report/42-196/196C.pdf}.

\bibitem{IAAEPM}
E.~V. Pitjeva and N.~P. Pitjev, ``Development of planetary ephemerides epm and
  their applications,'' \href{http://dx.doi.org/10.1007/s10569-014-9569-0}{{\em
  Celestial Mechanics and Dynamical Astronomy} {\bfseries 119} no.~3, (Aug,
  2014) 237--256}. \url{http://dx.doi.org/10.1007/s10569-014-9569-0}.

\bibitem{Tremaine:1979we}
S.~Tremaine and J.~E. Gunn, ``{Dynamical Role of Light Neutral Leptons in
  Cosmology},''
\href{http://dx.doi.org/10.1103/PhysRevLett.42.407}{{\em Phys. Rev. Lett.}
  {\bfseries 42} (1979) 407--410}.

\bibitem{Fischbach:1992fa}
E.~Fischbach and C.~Talmadge, ``{Six years of the fifth force},''
\href{http://dx.doi.org/10.1038/356207a0}{{\em Nature} {\bfseries 356} (1992)
  207--214}.

\bibitem{Kanemura:2010sh}
S.~Kanemura, S.~Matsumoto, T.~Nabeshima, and N.~Okada, ``{Can WIMP Dark Matter
  overcome the Nightmare Scenario?},''
  \href{http://dx.doi.org/10.1103/PhysRevD.82.055026}{{\em Phys. Rev.}
  {\bfseries D82} (2010) 055026},
\href{http://arxiv.org/abs/1005.5651}{{\ttfamily arXiv:1005.5651 [hep-ph]}}.

\bibitem{Jungman:1995df}
G.~Jungman, M.~Kamionkowski, and K.~Griest, ``{Supersymmetric dark matter},''
  \href{http://dx.doi.org/10.1016/0370-1573(95)00058-5}{{\em Phys. Rept.}
  {\bfseries 267} (1996) 195--373},
\href{http://arxiv.org/abs/hep-ph/9506380}{{\ttfamily arXiv:hep-ph/9506380
  [hep-ph]}}.

\bibitem{SakuraiQM}
J.~J. Sakurai, {\em Modern Quantum Mechanics, Revised Edition}.
\newblock Addison Wesley, 1~ed., 9, 1993.
\newblock \url{http://amazon.co.jp/o/ASIN/0201539292/}.

\bibitem{MessiahQM}
A.~Messiah, {\em Quantum Mechanics (Dover Books on Physics)}.
\newblock Dover Publications, 2, 2014.
\newblock \url{http://amazon.co.jp/o/ASIN/048678455X/}.

\bibitem{LandauLifshitz198101}
L.~D. Landau and L.~M. Lifshitz, {\em Quantum Mechanics, Third Edition:
  Non-Relativistic Theory (Volume 3)}.
\newblock Butterworth-Heinemann, 3~ed., 1, 1981.

\bibitem{doi:10.1063/1.1722537}
S.~I. Rubinow and T.~T. Wu, ``First correction to the geometric-optics
  scattering cross section from cylinders and spheres,''
  \href{http://dx.doi.org/10.1063/1.1722537}{{\em Journal of Applied Physics}
  {\bfseries 27} no.~9, (1956) 1032--1039},
  \href{http://arxiv.org/abs/http://dx.doi.org/10.1063/1.1722537}{{\ttfamily
  http://dx.doi.org/10.1063/1.1722537}}.
  \url{http://dx.doi.org/10.1063/1.1722537}.

\bibitem{Lewin:1995rx}
J.~D. Lewin and P.~F. Smith, ``{Review of mathematics, numerical factors, and
  corrections for dark matter experiments based on elastic nuclear recoil},''
\href{http://dx.doi.org/10.1016/S0927-6505(96)00047-3}{{\em Astropart. Phys.}
  {\bfseries 6} (1996) 87--112}.

\bibitem{Guth:2014hsa}
A.~H. Guth, M.~P. Hertzberg, and C.~Prescod-Weinstein, ``{Do Dark Matter Axions
  Form a Condensate with Long-Range Correlation?},''
  \href{http://dx.doi.org/10.1103/PhysRevD.92.103513}{{\em Phys. Rev.}
  {\bfseries D92} no.~10, (2015) 103513},
\href{http://arxiv.org/abs/1412.5930}{{\ttfamily arXiv:1412.5930
  [astro-ph.CO]}}.

\bibitem{Chakrabarty:2017fkd}
S.~S. Chakrabarty, S.~Enomoto, Y.~Han, P.~Sikivie, and E.~M. Todarello,
  ``{Gravitational self-interactions of a degenerate quantum scalar field},''
\href{http://arxiv.org/abs/1710.02195}{{\ttfamily arXiv:1710.02195 [hep-ph]}}.

\bibitem{Krasinsky2004}
G.~A. Krasinsky and V.~A. Brumberg, ``Secular increase of astronomical unit
  from analysis of the major planet motions, and its interpretation,''
  \href{http://dx.doi.org/10.1007/s10569-004-0633-z}{{\em Celestial Mechanics
  and Dynamical Astronomy} {\bfseries 90} no.~3, (Nov, 2004) 267--288}.
  \url{http://dx.doi.org/10.1007/s10569-004-0633-z}.

\bibitem{standish_2004}
E.~M. Standish, ``The astronomical unit now,''
  \href{http://dx.doi.org/10.1017/S1743921305001365}{{\em Proceedings of the
  International Astronomical Union} {\bfseries 2004} no.~IAUC196, (2004)
  163--179}.

\bibitem{Arvanitaki:2014wva}
A.~Arvanitaki, M.~Baryakhtar, and X.~Huang, ``{Discovering the QCD Axion with
  Black Holes and Gravitational Waves},''
  \href{http://dx.doi.org/10.1103/PhysRevD.91.084011}{{\em Phys. Rev.}
  {\bfseries D91} no.~8, (2015) 084011},
\href{http://arxiv.org/abs/1411.2263}{{\ttfamily arXiv:1411.2263 [hep-ph]}}.

\bibitem{Cardoso:2018tly}
V.~Cardoso, O.~J.~C. Dias, G.~S. Hartnett, M.~Middleton, P.~Pani, and J.~E.
  Santos, ``{Constraining the mass of dark photons and axion-like particles
  through black-hole superradiance},''
\href{http://arxiv.org/abs/1801.01420}{{\ttfamily arXiv:1801.01420 [gr-qc]}}.

\bibitem{MukaidaYanagida}
T.~T. Yanagida and K.~Mukaida {\em In preparation} .

\bibitem{Hagmann:1999kf}
C.~Hagmann, ``{Cosmic neutrinos and their detection},'' in {\em {American
  Physical Society (APS) Meeting of the Division of Particles and Fields (DPF
  99) Los Angeles, California, January 5-9, 1999}}.
\newblock 1999.
\newblock
\href{http://arxiv.org/abs/astro-ph/9905258}{{\ttfamily arXiv:astro-ph/9905258
  [astro-ph]}}.
\newblock

\bibitem{Seoane:2013qna}
{\bfseries eLISA} Collaboration, P.~A. Seoane {\em et~al.}, ``{The
  Gravitational Universe},''
\href{http://arxiv.org/abs/1305.5720}{{\ttfamily arXiv:1305.5720
  [astro-ph.CO]}}.

\bibitem{Kawasaki:1991eu}
M.~Kawasaki, H.~Murayama, and T.~Yanagida, ``{Can the strongly interacting dark
  matter be a heating source of Jupiter?},''
\href{http://dx.doi.org/10.1143/PTP.87.685}{{\em Prog. Theor. Phys.} {\bfseries
  87} (1992) 685--692}.

\end{thebibliography}\endgroup
\end{document}